\documentclass[a4paper,11pt]{article}
\pdfoutput=1 

\usepackage{jheppub} 

\usepackage[T1]{fontenc} 
\usepackage{orcidlink }
\usepackage{tikz}
\usepackage{orcidlink}
\usetikzlibrary{decorations.pathmorphing}
\usepackage{graphicx}
\graphicspath{ {./images/} }
\usepackage[toc,page]{appendix}

\title{\boldmath A timelike entangled island at the initial singularity in a JT FLRW ($\Lambda>0$) universe}

\author[1]{K. Sreeman Reddy\,\orcidlink{0000-0002-9897-9573}}
\affiliation[1]{Department of Physics, Indian Institute of Technology Bombay,\\Powai, Mumbai, 400076, India}

\emailAdd{sreeman@iitb.ac.in}

\abstract{It has been argued that there are no islands in FLRW cosmologies with $\Lambda>0$ and $k=0$ \cite{Hartman:2020khs}. We argue that there is a timelike separated island at the initial singularity, and it will resolve the cosmological information paradox. The information about the particles that went beyond the horizon is not lost for our observer. By measuring Hawking radiation, we can get that information from the past when those particles were near the initial singularity. Similar to how islands inside black holes violate locality, we observe a violation of causality or \textit{noncausality} but \textit{only at the initial singularity}, possibly the only region where it is \textit{acceptable}. We start with a review of timelike entanglement. We will follow an approach similar to the one followed in \cite{Aalsma:2021bit} for normal islands. In the end, we conjecture a generalization of the Ryu–Takayanagi or QES prescription for the case of bulk timelike entanglement in dS/CFT correspondence and comment on the emergence of time in dS/CFT correspondence.}

\begin{document} 
\addtocontents{toc}{\protect{\pdfbookmark[1]{\contentsname}{toc}}}
\maketitle
\flushbottom

\section{Introduction}
\label{sec:intro}

In 1916, even before the advent of modern quantum mechanics, Einstein realized that his general relativity needs to be modified to make it compatible with quantum mechanics \cite{Einstein:1916cc}. More than a century later, quantum gravity is still not understood. String theory \cite{Kiritsis:2019npv} is so far the most successful attempt for a theory of quantum gravity. Using semiclassical gravity, Hawking argued that black holes \textit{do} radiate \cite{Hawking:1975vcx} and that information that falls into them is permanently lost \cite{Hawking:1976ra}. This is at serious odds with the unitarity of quantum physics and is called the \textit{black hole information paradox}\footnote{For a review of black hole information paradox check \cite{Harlow:2014yka}.}.

In the last 25 years, the AdS/CFT correspondence \cite{Maldacena:1997re,Witten:1998qj,Gubser:1998bc,Aharony:1999ti}, a relatively well-understood example of the holographic principle \cite{tHooft:1993dmi,Susskind:1994vu}, has greatly advanced our understanding of both non-perturbative quantum gravity and strongly coupled quantum field theories\footnote{For introductory reviews of the AdS/CFT correspondence check \cite{Nastase:2015wjb,Ammon:2015wua,Natsuume:2014sfa}.}. In particular, it answered that black hole evaporation is unitary and information is not lost albeit initially it was not known how exactly this happens. Ryu and Takayanagi argued that the \textit{entanglement entropy}\footnote{Actually it is the \textit{von Neumann entropy} and only when the total system is pure we can call it as the \textit{entanglement entropy}. But, this abuse of terminology is ubiquitous as we mostly deal with pure state total systems.} of a boundary subregion is given by a minimal-area surface in the bulk spacetime \cite{Ryu:2006bv,Ryu:2006ef}. Later this RT formula was generalized to the covariant HRT formula \cite{Hubeny:2007xt}. It was further generalized, to the regime where the bulk quantum corrections are not negligible by replacing classical extremal surfaces with \textit{quantum extremal surfaces} \cite{Faulkner:2013ana,Engelhardt:2014gca}\footnote{For reviews of the quantum information perspective of holography check \cite{Headrick:2019eth,Nishioka:2009un,VanRaamsdonk:2016exw,Rangamani:2016dms,Harlow:2018fse}.}.

Recently the black hole information paradox has been resolved \cite{Penington:2019npb,Almheiri:2019psf,Almheiri:2019hni,Penington:2019kki,Almheiri:2019qdq}. Even though initially everyone expected that a full quantum gravity theory is needed to resolve the information paradox, later it was realised that since the paradox starts at the Page time, when the curvature is still much larger than the Planck length, it must be solvable within semiclassical gravity. After solving the information paradox using holography \cite{Penington:2019npb,Almheiri:2019psf,Almheiri:2019hni}, it was soon generalized to semiclassical gravity using replica wormholes in the gravitational path integral \cite{Penington:2019kki,Almheiri:2019qdq}. An important concept called \textbf{island} was the key to resolving the paradox. These islands are formed in the interior of the black hole after Page time, and are connected \textit{nonlocally} to the outside radiation. By giving up locality, the island prescription resolves the AMPS firewall paradox \cite{Almheiri:2012rt}. Most of this work was done initially in low-dimensional models and was later generalized to higher dimensions\footnote{For an intuitive review check \cite{Almheiri:2020cfm} and for a proper review check \cite{Mertens:2022irh}.}.

Compared to AdS holography, dS holography \cite{Strominger:2001pn,Anninos:2011ui} is poorly understood. Similar to the AdS case\footnote{For a review of AdS$_{2}$ holography and the SYK model check \cite{Sarosi:2017ykf,Trunin:2020vwy}.}, it will be easier to study lower dimensional cases \cite{Maldacena:2019cbz,Cotler:2019nbi} before going to the higher dimensional cases. Recently islands have been studied in dS spacetimes \cite{Aalsma:2021bit,Chen:2020tes,Hartman:2020khs,Balasubramanian:2020xqf,Sybesma:2020fxg,Kames-King:2021etp,Bousso:2022gth,Espindola:2022fqb,Levine:2022wos,Aalsma:2022swk}. Some argued that since islands are needed to solve the cosmological information paradox, the universe has a small (as yet undetected) non-zero spatial curvature ($k\neq 0$), and this will allow the existence of islands in our universe \cite{Bousso:2022gth,Espindola:2022fqb}. But here, we argue that $k\neq 0$ is not needed, and an island can exist even in $k= 0$ once we include timelike entanglement.

\section{Timelike entanglement}

The violation of the CHSH inequality \cite{Clauser:1969ny} implies that nature cannot be explained within local hidden-variable theories. Even though EPR \cite{Einstein:1935rr} famously got confused that entanglement is a nonlocal phenomenon, in the EPR pair, we cannot influence the measurement of the particle near us. So, we cannot instantaneously send information to the particle that is far away. Entanglement is just a \textit{stronger form of correlation} between particles than what is possible in classical physics.

The violation of the Leggett–Garg inequality \cite{Leggett:1985zz} implies that the \textit{time evolution of a quantum system cannot be understood classically}. This implies that \textbf{timelike entanglement} is necessarily present in any quantum theory. Although it is present in all quantum theories very few papers related to timelike entanglement have been written \cite{Olson:2010jy,Olson:2011bq,Sabin:2012pj}\footnote{For a review check section 6.4 of \cite{Rajan:2020qoy}.}. Timelike entanglement is not a crazy speculative idea; it is even \textbf{experimentally verified} \cite{Megidish_2013}.

Timelike entanglement does not violate causality just like normal (or spacelike) entanglement does not violate locality\footnote{We are following the High Energy Physics terminology here. In information theory terminology, they often call normal entanglement a nonlocal phenomenon even though it does not allow for faster-than-light communication.}. It is just a \textit{stronger form of correlation} between past and future than what is possible in classical physics. If two particles are timelike entangled, then \textit{we cannot influence the particle in the past} by measuring the particle in the future since \textit{we cannot control the measurement result} of the future particle.

\textbf{The general definition of (spacelike) entanglement}: Consider a spacetime manifold $M$. Its Hilbert space is defined on a Cauchy slice $\Sigma$, which can be divided into two \textit{disjoint} subregions, $\Sigma=A \sqcup A^c$. Assume that the Hilbert space on $\Sigma$ can be factorized into $\mathcal{H}_{\Sigma}=\mathcal{H}_A \otimes \mathcal{H}_{A^c}$. Then the density matrix of the quantum field $\rho$ on the Cauchy slice $\Sigma$ is called \textbf{entangled} if it is not separable, meaning if it cannot be represented as
$$
\rho=\sum_i p_i \rho_{A}^i \otimes \rho_{A^c}^i
$$
where $\rho_{A}^i$ are density operators on $\mathcal{H}_{A}$ and $\rho_{A^c}^i$ are density operators on $\mathcal{H}_{A^c}$, with $p_i \geq 0$.

Then the von Neumann entropy is defined in terms of the reduced density matrix $\rho_A$ as
$$
S_{\mathrm{vN}}\left(\rho_A\right):=-\operatorname{Tr}\left(\rho_A \ln \rho_A\right)
$$
If the total state on $\Sigma$ is pure, then the von Neumann entropy measures the entanglement between the region $A$ and $A^c$. However, there is no good measure for the amount of entanglement between $A$ and $B \subset A^c$. In fact, there is no simple way other than to use the separable definition to even say if there is entanglement between $A$ and $B$ since together $A\cup B$ might be a mixed state (also true if the total state on $\Sigma$ is mixed). This is called the quantum separability problem \cite{Ioannou}.

The entanglement entropy is actually not a property of the particular Cauchy slice or subregion. It is actually a property of the \textit{domain of dependence} of the subregion. For example, in Fig \ref{fig:compare} the two different subregions (shown in black) in the left blue region have the same domain of dependence (left blue region). Now we will generalize entanglement entropy so that timelike entanglement entropy is a subset of it.

\subsection{Definition of general entanglement}
Consider a spacetime manifold $M$. Take 2 Cauchy slices $\Sigma_1$ \& $\Sigma_2$. Now take two \textit{disjoint} subregions of both $\Sigma_1$ \& $\Sigma_2$ such that the \textit{intersection of their domain of dependences is a null set}. Let them be $A\subset \Sigma_1$ and $B\subset \Sigma_2$ so that $D_A \cap D_B=\emptyset$. Assume that the Hilbert space of the fields when \textbf{restricted} to $D_A \cup D_B$ can be factorized into $\mathcal{H}=\mathcal{H}_{D_A} \otimes \mathcal{H}_{D_B}$. Then the density matrix of the quantum field $\rho$ restricted to $D_A \cup D_B$ is called \textbf{entangled} if it is not separable, meaning if it cannot be represented as
$$
\rho=\sum_i p_i \rho_{D_A}^i \otimes \rho_{D_B}^i
$$
where $\rho_{D_A}^i$ are density operators on $\mathcal{H}_{D_A}$ and $\rho_{D_B}^i$ are density operators on $\mathcal{H}_{D_B}$, with $p_i \geq 0$.

In the previous definition of (spacelike) entanglement, we did not need to specify that the intersection of their domain of dependences is a null set because it is \textit{always trivially true} when we take subregions of the same Cauchy slice.

\begin{figure}[h]
\includegraphics[width=14cm]{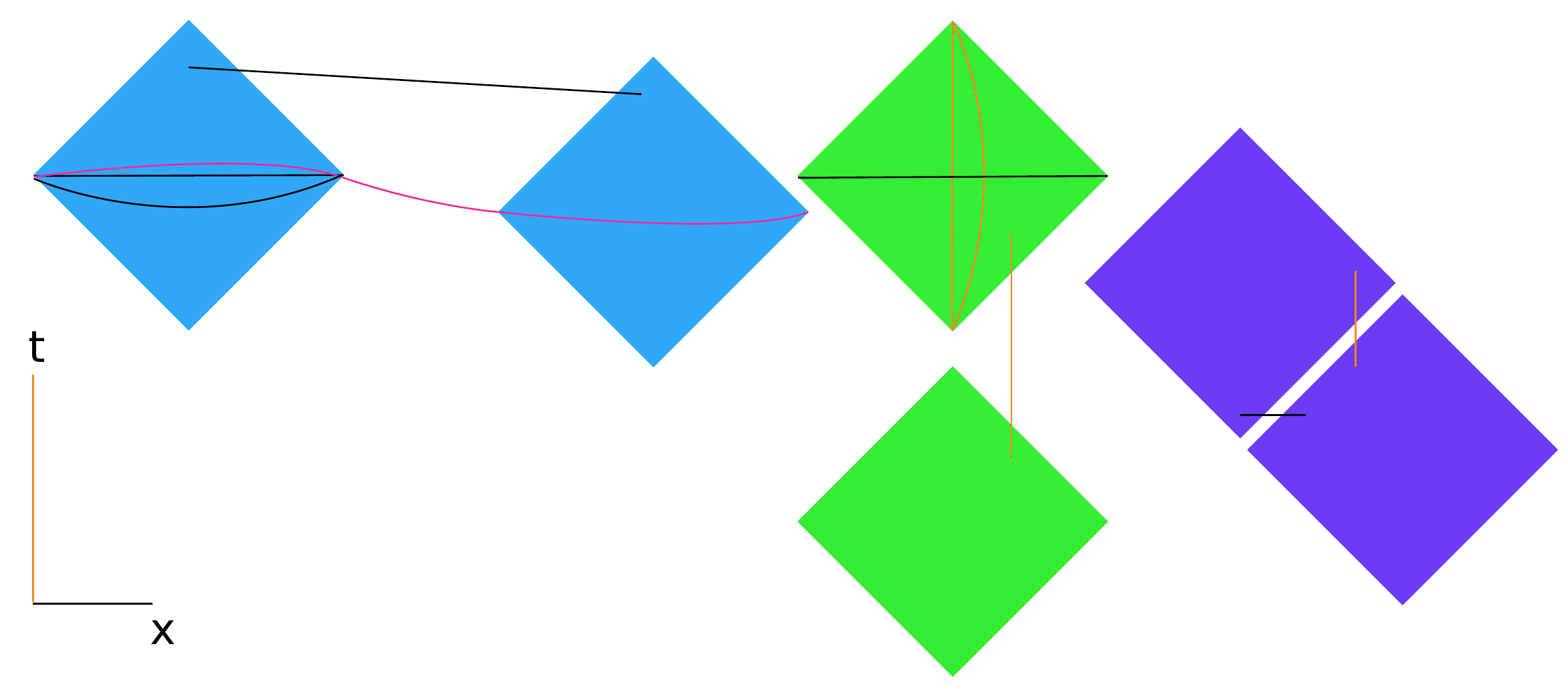}
\centering
\caption{Timelike separations are given in orange, and spacelike separations are given in black. The separation between any 2 points from the two blue regions is spacelike. The 2 black subregions define the same domain of dependence (left blue region). So, they are spacelike entangled, and we can find a Cauchy slice (e.g. pink slice) that contains 2 subregions whose domain of dependences are these blue regions. The entanglement between \textbf{green} regions is \textbf{timelike}, whereas between the purple regions is mixed.}
\label{fig:compare}
\end{figure}

Essentially, if two domain of dependences are such that all points between them are timelike separated then the entanglement between them is \textbf{timelike entanglement}. Similarly, \textbf{spacelike entanglement} and \textbf{mixed entanglement} are shown in Fig \ref{fig:compare}. When the two domain of dependences are such that some points between them are spacelike whereas some are timelike separated then the entanglement between them is mixed entanglement. Whenever the two domain of dependences are such that all points between them are spacelike separated then we can always find a Cauchy slice that contains 2 subregions whose domain of dependences are these domain of dependences as shown in pink in Fig \ref{fig:compare}.

We can define \textbf{timelike entanglement entropy} for the case of timelike entanglement using path integrals to calculate the density matrix as will be clear when we discuss the timelike entanglement entropy for an interval in $2D$ CFTs. For timelike entanglement also we will not have a good measure for the entanglement between 2 random domain of dependences $D_A$ and $D_B$. The only case where we can measure the entanglement is between a domain of dependence and its \textit{some type of compliment}. The union of the future of $D_A$ (i.e. set all points that are in the future cone of \textit{every} point in $D_A$) and the past of $D_A$ (i.e. set all points that are in the past cone of \textit{every} point in $D_A$) is the \textit{natural analog} to the $D_{A^c}$ in the spacelike entanglement case.

An important difference between timelike and spacelike entanglement is that if we consider a Cauchy slice and find the total spacelike entanglement entropy of the Cauchy slice, it will be constant throughout time (in general for simplicity we assume the constant as $0$ i.e. we assume that the state is pure on a Cauchy slice). This difference is a \textit{direct consequence} of the fact that there is a specific \textit{arrow of time}. Since even if we know everything about a co-dimension 1 Lorentzian slice (unlike Cauchy slices, which are Euclidean), we still do not know everything about the entire spacetime because there are \textit{no arrows for spatial dimensions}.

\begin{figure}[h]
\includegraphics[width=6cm]{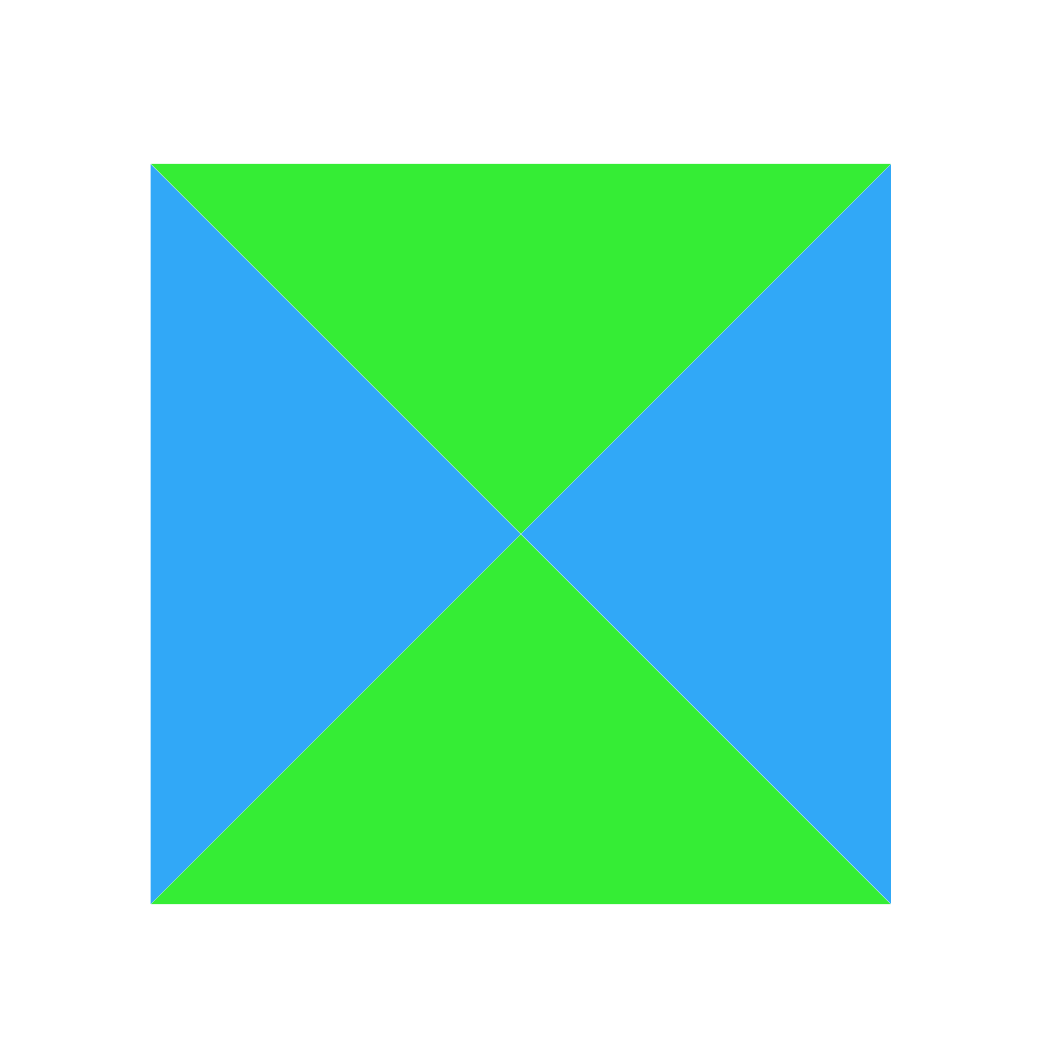}
\centering
\caption{The future and past light cones are in green and the Rindler wedges are in blue. The entanglement entropy between same colour regions is equal.}
\label{fig:FP}
\end{figure}

In \cite{Olson:2010jy}, it was shown that the entanglement between future and past light cones of a scalar field is exactly the same as the entanglement between the left and right Rindler wedges, see Fig \ref{fig:FP}. That is the usual spacelike entanglement present in the Unruh effect given by
$$\left|0_M\right\rangle=\prod_i C_i \sum_{n_i=0}^{\infty} \frac{e^{-\pi n_i \omega_i / a}}{n_{i} !}\left(\hat{a}_{\omega_i}^{R_i^{\dagger}} \hat{a}_{\omega_i}^{L \dagger}\right)^{n_i}\left|0_R\right\rangle, $$
$$\hat{\rho}_R=\prod_i\left[C_i^2 \sum_{n_i=0}^{\infty} e^{-2 \pi n_i \omega_i / a}\left|n_i^R\right\rangle\left\langle n_i^R\right|\right],$$
is generalised to
$$\left|0_M\right\rangle=\prod_i C_i \sum_{n_i=0}^{\infty} \frac{e^{-\pi n_i \omega_i / a}}{n_{i} !}\left(\hat{a}_{\omega_i}^{F \dagger} \hat{a}_{\omega_i}^{P \dagger}\right)^{n_i}\left|0_T\right\rangle,$$

$$\hat{\rho}_F=\prod_i\left[C_i^2 \sum_{n_i=0}^{\infty} e^{-2 \pi n_i \omega_i / a}\left|n_i^F\right\rangle\left\langle n_i^F\right|\right].$$

\subsection{Timelike entanglement entropy for an interval in $2D$ CFTs}

In $2D$ CFTs, there is a universal formula for the entanglement entropy of an interval. If we take an interval with length $L$, then the entropy of that interval is
$$
S=\frac{c}{3} \ln \frac{L}{\epsilon}
$$
The generalization of this formula\footnote{Originally I gave intuitive reasons for this formula and conjectured it. But near the completion of this paper I noticed \cite{Doi:2022iyj} where they derived it.} for the case of a \textbf{timelike interval} with length $T$ is obtained by replacing $L$ with $iT$ \cite{Doi:2022iyj} since $ds^2$ is $-$ve for timelike separation.
\begin{equation}\label{eq:timelikeinterval}
S=\frac{c}{3} \ln \frac{iT}{\epsilon}=\frac{c}{3} \ln \frac{T}{\epsilon}+\frac{c}{3}i(\frac{\pi}{2}+2k\pi)
\end{equation}
Since the imaginary part is not unique, it is not physically relevant, and we will discard it.
\begin{figure}[h]
\includegraphics[width=8cm]{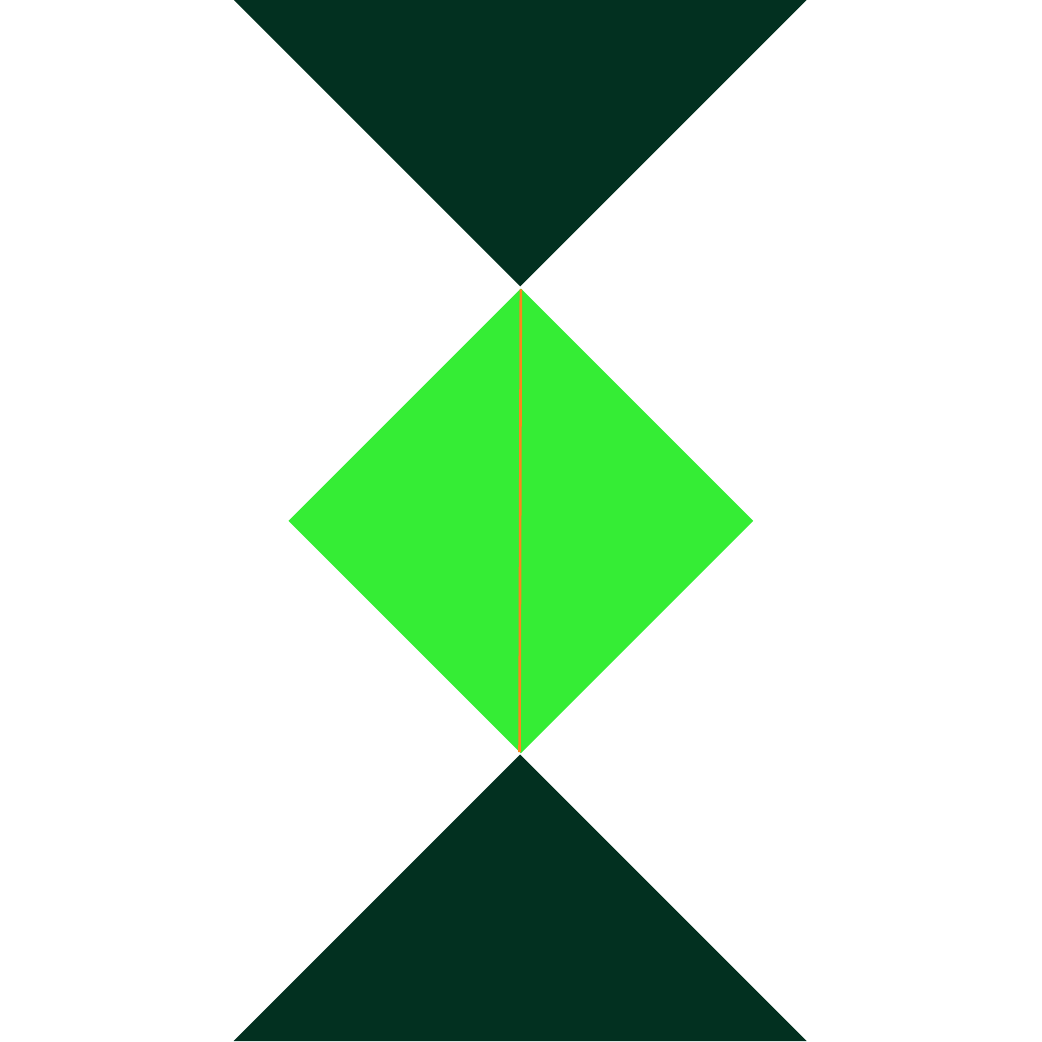}
\centering
\caption{The timelike entanglement entropy formula for an interval gives the entanglement between the light green region and the dark green region. The dark green region is the union of the past and the future of the domain of dependence given here. The length of the interval shown in orange gives the entropy. Notice that this is just a $90^{\circ}$ rotation of the spacelike interval case.}
\label{fig:2dcft}
\end{figure}

If we go to a coordinate system where the metric is in a conformal gauge
$$\mathrm{d} s^2=-\Omega^{-2} \mathrm{~d} y^{+} \mathrm{d} y^{-},$$
the entanglement entropy of an interval between $(y^{+}_1,y^{-}_1)$ and $(y^{+}_2,y^{-}_2)$ in the \textit{vacuum state} is
\begin{equation}\label{eq:vne}
S_{\mathrm{vN}}\left(y_1, y_2\right)=\frac{c}{12} \log \left[\frac{\left(y_1^{+}-y_2^{+}\right)^2}{\epsilon^2 \Omega\left(y_1\right) \Omega\left(y_2\right)}\right]+\frac{c}{12} \log \left[\frac{\left(y_1^{-}-y_2^{-}\right)^2}{\epsilon^2 \Omega\left(y_1\right) \Omega\left(y_2\right)}\right].
\end{equation}
Notice that this formula is of the form $S=\frac{c}{12} \ln \frac{L^4}{\epsilon^4}$. So, this formula is also valid for the timelike case, and the imaginary part is \textit{automatically discarded} since $(iT)^4=T^4$.

The formula can be proved for the spacelike case using the replica trick \cite{Headrick:2019eth,VanRaamsdonk:2016exw}. The density matrix can be calculated by the path integral on the Euclidean plane cut along the interval on the $x$-axis. Then to get the trace of the $n$th power of the density matrix, we have to take $n$ copies of the manifold and cyclically attach the cuts to one another. We can then calculate the $n$th Rényi entropies and take the limit $n\to 1$ to get the von Neumann entropy.

The derivation for the timelike case is also similarly done in \cite{Doi:2022iyj}. We need to calculate the path integral on the Euclidean plane cut along the time interval on the time axis.

\subsubsection{$2D$ is special}

Many things are unique to $2D$ dimensional spacetimes. In $2D$, connected spatial subregions are always intervals. If we add one more spatial dimension (i.e., $(2+1)D$ or $3D$ spacetimes), then we can have many different kinds of spatial subregions like disks, squares, triangles, etc.

In $2D$, a \textit{timelike interval} uniquely determines a domain of dependence because the number of spacelike and timelike dimensions are equal, as shown in Fig \ref{fig:compare} where the green region can be defined both by the black interval or the orange intervals. In higher dimensions, this is not the case. For example, in $3D$ spacetimes, the domain of dependences of a square and disk can have the same maximum time interval.

The \textit{universality} of the spacelike entanglement entropy of an interval in $2D$ CFT is well known. Similarly, for an interval in $2D$ CFT, the timelike entanglement entropy is universal and depends only on the length of the interval.

\section{FLRW universe in JT gravity}
Einstein's general relativity is topological in $1+1$ dimensions. But, in JT gravity \cite{Mertens:2022irh,Jackiw:1984je,Teitelboim:1983ux}, because of the dilaton, we can see interesting dynamics. The action of de Sitter JT gravity with conformal matter is
\begin{align*}
    I&=\frac{\Phi_0}{2 \pi}\left(\int \mathrm{d}^2 x \sqrt{-g} R-2 \int \mathrm{d} x \sqrt{|h|} K\right) \\
&+\frac{1}{2 \pi}\left(\int \mathrm{d}^2 x \sqrt{-g} \Phi\left(R-\frac{2}{\ell^2}\right)-2 \int \mathrm{d} x \sqrt{|h|} \Phi K\right)+I_{\mathrm{CFT}}.
\end{align*}

Here the first term is the topological Einstein-Hilbert term, second term is the JT term, and the last term is the matter term. Now from the Euler–Lagrange equations, the equations of motion for the dilaton and the metric, respectively, in the semiclassical regime are

$$\begin{gathered}
R-2 / \ell^2=0, \\
\Phi g_{\mu\nu}-\ell^2 \nabla_\mu \nabla_\nu \Phi+\ell^2 g_{\mu \nu} \square \Phi=\pi \ell^2\left\langle T_{\mu \nu}\right\rangle .
\end{gathered}$$
For simplicity, we will fix the length scale as $\ell=1$ from now on. Here $\left\langle T_{\mu \nu}\right\rangle$ is the expectation value of the matter quantum fields. Notice that \textit{all the backreaction} caused by the introduction of the matter is contained in the dilaton field and that the metric remains unchanged.

We will, from now on, focus on the following Milne solution to the above equations.

$$d s_{\text {Milne }}^2=-d t^2+\sinh^2(t) d x^2, \quad \Phi=\frac{\Phi_s}{24} \cosh (t)$$
This is an FLRW universe with the scale factor $a(t)=\sinh(t)$. The spatial curvature is \textit{trivially} $k=0$ since 1D spatial lines do not have intrinsic curvature. The $24$ factor is introduced for future simplifications. We can now define comoving spacelike coordinate $X$ and null coordinates $(x^+,x^-)$ and static null coordinates $(\sigma^+,\sigma^-)$ as

$$X=a(t)x=\sinh(t)x$$
$$x^+=\tanh(t/2)e^{+x},\quad x^-=\tanh(t/2)e^{-x}$$
$$\sigma^+=\ln{x^+},\quad \sigma^-=\ln{x^-}$$

These coordinate systems then give

$$d s^2=-(1-H^2(t)X^2)d t^2-2H(t)XdXdt+d X^2, \quad \Phi=\frac{\Phi_s}{24} \cosh (t)$$

$$
\mathrm{d} s^2=-\frac{4}{\left(1-x^{+} x^{-}\right)^2} \mathrm{~d} x^{+} \mathrm{d} x^{-}, \quad \Phi=\frac{\Phi_s}{24} \left(\dfrac{1+x^{+}x^{-}}{1-x^{+}x^{-}}\right)
$$

$$
\mathrm{d} s^2=-\text{cosech}^2\left(\frac{\sigma^{+}+\sigma^{-}}{2}\right) \mathrm{~d} \sigma^{+} \mathrm{d} \sigma^{-}, \quad \Phi=-\frac{\Phi_s}{24} \coth \left(\frac{\sigma^{+}+\sigma^{-}}{2}\right)
$$
Here $H(t)=\frac{\Dot{a}}{a}=\coth(t)$. We can maximally extend this spacetime to get a solution that also contains a black hole, as was considered in \cite{Aalsma:2021bit,Maldacena:2019cbz} but doing this is not necessary since $t=0$ is a singularity\footnote{Some might argue that this is not a physical singularity, but a coordinate singularity in $(t,x)$ coordinates where $t=0$ is a point, unlike in $(t,X)$ where it is a $3D$ surface. However, if we calculate \textbf{matter density (per unit length)} for some classical dust, it \textbf{diverges} at $t=0$. We will soon see (Eq \ref{eq:T}) that even in our semiclassical model with conformal matter, we see divergence at $t=0$, so it is a singularity. But, there is no good definition of a singularity \cite{Geroch:1968ut}, so this discussion will never conclude.}.

We can see that for an observer at the origin, there are 2 \textit{disconnected}\footnote{This is an important difference compared to higher dimensions. This is analogous to how a circle in $2D$ space reduces to 2 disconnected points in $1D$ space when we make a dimensional reduction. In higher dimensions, a particle emitted by one side of the horizon can be absorbed by the opposite side of the horizon.} horizons at

$$X_h=\pm \dfrac{1}{\sqrt{H}}=\pm \sqrt{tanh(t)},\quad x_h=\pm \sqrt{\dfrac{2}{\sinh (2t)}}$$

\begin{figure}[h]
\includegraphics[width=8cm]{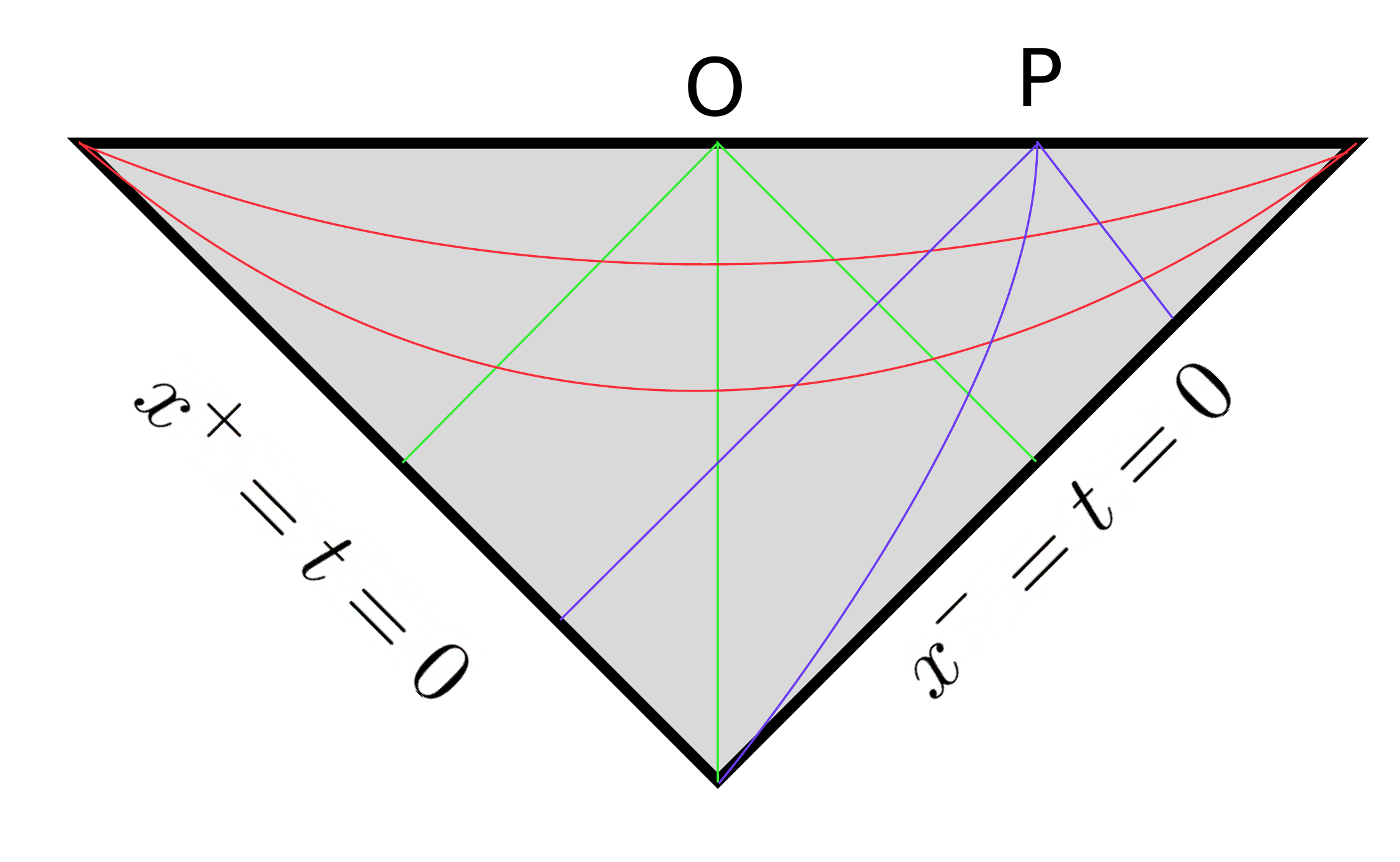}
\centering
\caption{Penrose diagram of our 2D FLRW spacetime. The horizons of the observer O (green) with $X=0$ and the observer P (purple) with some $X>0$ are different. The red lines are constant $t$ slices.}
\label{fig:FLRW}
\end{figure}

For $t>2$, $X_h\approx\pm 1$. These horizons are observer dependent as shown in Figure \ref{fig:FLRW}.

\subsection{Conformal matter and its backreaction}

We now include conformal matter and calculate its backreaction on the dilaton. Doing this is preferable in the null coordinates $(x^+,x^-)$.

In $2D$ CFT, the trace of the stress-energy-momentum tensor is entirely fixed by the conformal or Weyl anomaly for all states and is given by \cite{Kiritsis:2019npv}

$$\left\langle T_a^a\right\rangle=\frac{c}{12 \pi \ell^2}=\frac{c}{12 \pi}$$
Our metric in $(x^+,x^-)$ completely is off-diagonal, so the off-diagonal terms cannot be zero even in the ground state.

We now focus on the following state called the Bunch-Davies vacuum state, defined by

$$
\left\langle T_{\pm \pm}\left(x^{\pm}\right)\right\rangle_{\mathrm{BD}}=0 .
$$
For the sake of intuition and to understand what a static observer would see, we can transform to the static null coordinates $(\sigma^+,\sigma^-)$. Because of the conformal anomaly, the stress–energy–momentum tensor does not transform as a tensor and instead follows
\begin{equation}\label{eq:TT}
    \left\langle T_{\pm \pm}\left(\sigma^{\pm}\right)\right\rangle=\left(x^{\pm \prime}\right)^2\left\langle T_{\pm \pm}\left(x^{\pm}\right)\right\rangle-\frac{c}{24 \pi}\left\{x^{\pm}, \sigma^{\pm}\right\}
\end{equation}

where the second term is the Schwarzian derivative, and $'$ denotes differentiation with respect to $\sigma^{\pm}$.
$$
\left\{x^{\pm}, \sigma^{\pm}\right\}=\frac{x^{\pm^{\prime \prime \prime}}}{x^{\pm^{\prime}}}-\frac{3}{2}\left(\frac{x^{\pm^{\prime \prime}}}{x^{\pm^{\prime}}}\right)^2
$$
Plugging $x^{\pm}=e^{\sigma^\pm}$ we get
$$\left\langle T_{\pm \pm}\left(\sigma^{\pm}\right)\right\rangle_{\mathrm{BD}}=\frac{c}{48 \pi \ell^2}=\frac{c}{48 \pi} $$
In these coordinates, it is clear that this state is a thermal equilibrium state where the amount of radiation moving to the right and moving to the left are the same. Both have the same inverse temperature $\beta=2 \pi \ell=2\pi$.

Using the equation of motion for the dilaton, we can get the following backreacted solution

$$\Phi(x^{+},x^{-})=\frac{1}{24}\left(c+\Phi_s \left(\dfrac{1+x^{+}x^{-}}{1-x^{+}x^{-}}\right)\right),$$

This solution is still independent of $t$ and the only difference compared to without conformal matter is that the dilaton is shifted by $\frac{c}{24}$.

\subsubsection{Non-equilibrium vacuum state}
To observe non-trivial dynamics, we define the following non-equilibrium vacuum state
$$
\left\langle T_{\pm \pm}\left(\sigma_{\pm}\right)\right\rangle=\frac{\pi c}{12 \beta_{\pm}^2},
$$
where $\beta_{-}$ and $\beta_{+}$ are the inverse temperatures of the right and left moving radiation, respectively.

We consider the case where $\beta^{-}=2\pi$ but $\beta^{+}>>1$; that is, the temperature of the left-moving radiation is very low. In this state, very little radiation goes to the left horizon, and almost all the radiation goes to the right horizon. So, in this state, the \textbf{left horizon shrinks} and \textbf{evaporates}, whereas the \textbf{right horizon keeps growing}. This is shown in Fig \ref{fig:island}.

Let us define
$$t_{\pm}=\frac{2 \pi}{\beta_{\pm}} \implies  t_{-}=1 \,\text{ and }\, t_{+}\approx 0.$$
If we now go back to the null coordinates $(x^+,x^-)$ by using \ref{eq:TT}, the stress–energy–momentum tensor becomes
\begin{align}\label{eq:T}
    \left\langle T_{++}\left(x^{+}\right)\right\rangle&=-\frac{c}{48 \pi\left(x^{+}\right)^2}\left(1-t_{+}^2\right)\approx-\frac{c}{48 \pi\left(x^{+}\right)^2},\\
    \left\langle T_{--}\left(x^{-}\right)\right\rangle&=0.
\end{align}

Now the back reaction caused by the above stress–energy–momentum tensor gives the following solution for the dilaton equations of motion
\begin{align}
    \Phi\left(x^{+}, x^{-}\right)&=\frac{c}{48}\left[1+\frac{2 \Phi_s}{c} \left(\dfrac{1+x^{+}x^{-}}{1-x^{+}x^{-}}\right)+t_{+}^2-\left(1-t_{+}^2\right) \left(\dfrac{1+x^{+}x^{-}}{1-x^{+}x^{-}}\right) \log \left(x^{+}\right)\right]\\\label{eq:Phi}
    &\approx \frac{c}{48}\left[1+\frac{2 \Phi_s}{c} \left(\dfrac{1+x^{+}x^{-}}{1-x^{+}x^{-}}\right)-\left(\dfrac{1+x^{+}x^{-}}{1-x^{+}x^{-}}\right) \log \left(x^{+}\right)\right].
\end{align}

\section{Island}

We now calculate the page curve by using the island prescription. According to the island prescription, the full ﬁne-grained entropy of the radiation is given by

$$S(R)=\min \operatorname{ext}_I\left[2 \Phi_H(\partial I)+S_{\mathrm{vN}}(R \cup I)\right].$$

Here $2 \Phi_H(\partial I)$ is the 2D version of the $\dfrac{\text{Area}(\partial I)}{4G_N}$. Note that the LHS is valid even in a full quantum gravity theory since it is fine-grained, whereas the RHS is a semi-classical description. Usually, this formula is known to be valid only for the case of spacelike entanglement. Here we \textit{conjecture} that it is valid even in the case of timelike entanglement.

One solution to the island is the trivial case of the vanishing island. The von Neumann entropy can be calculated from the formula \ref{eq:vne}.

$$S_{\mathrm{vN}}\left(y_1, y_2\right)=\frac{c}{12} \log \left[\frac{\left(y_1^{+}-y_2^{+}\right)^2}{\epsilon^2 \Omega\left(y_1\right) \Omega\left(y_2\right)}\right]+\frac{c}{12} \log \left[\frac{\left(y_1^{-}-y_2^{-}\right)^2}{\epsilon^2 \Omega\left(y_1\right) \Omega\left(y_2\right)}\right].$$

However, since our non-equilibrium state is not a vacuum state in $(x^+,x^-)$, by using \ref{eq:TT} we search for $(y^+,y^-)$ such that it becomes a vacuum state with zero diagonal terms. We find that
$$y^{-}\left(x^{-}\right)=x^{-} \text{ and } y^{+}\left(x^{+}\right)= \log \left(x^{+}\right).$$
In these coordinates, the metric is
$$
\mathrm{d} s^2=-\frac{4 x^{+}}{\left(1-x^{+} x^{-}\right)^2} \mathrm{~d} y^{+} \mathrm{d} y^{-},
$$
with
$$\Omega^{-2}=\frac{4 x^{+}}{\left(1-x^{+} x^{-}\right)^2}. $$
So, the von Neumann entropy between points $A$ and $A'$ is

$$
\begin{aligned}
S_{\mathrm{vN}}(R) &=\frac{c}{12} \log \left[\frac{4 \left(x_A^{+} x_{A^{\prime}}^{+}\right)^{1 / 2}\left(x_A^{-}-x_{A^{\prime}}^{-}\right)^2}{\epsilon^2\left(1-x_A^{+} x_A^{-}\right)\left(1-x_{A^{\prime}}^{+} x_{A^{\prime}}^{-}\right)}\right] \\
&+\frac{c}{12} \log \left[\frac{4\left(x_A^{+} x_{A^{\prime}}^{+}\right)^{1 / 2} \log \left(x_{A^{\prime}}^{+} / x_A^{+}\right)^2}{\epsilon^2\left(1-x_A^{+} x_A^{-}\right)\left(1-x_{A^{\prime}}^{+} x_{A^{\prime}}^{-}\right)}\right] .
\end{aligned}
$$

\begin{figure}[h]
\includegraphics[width=14cm]{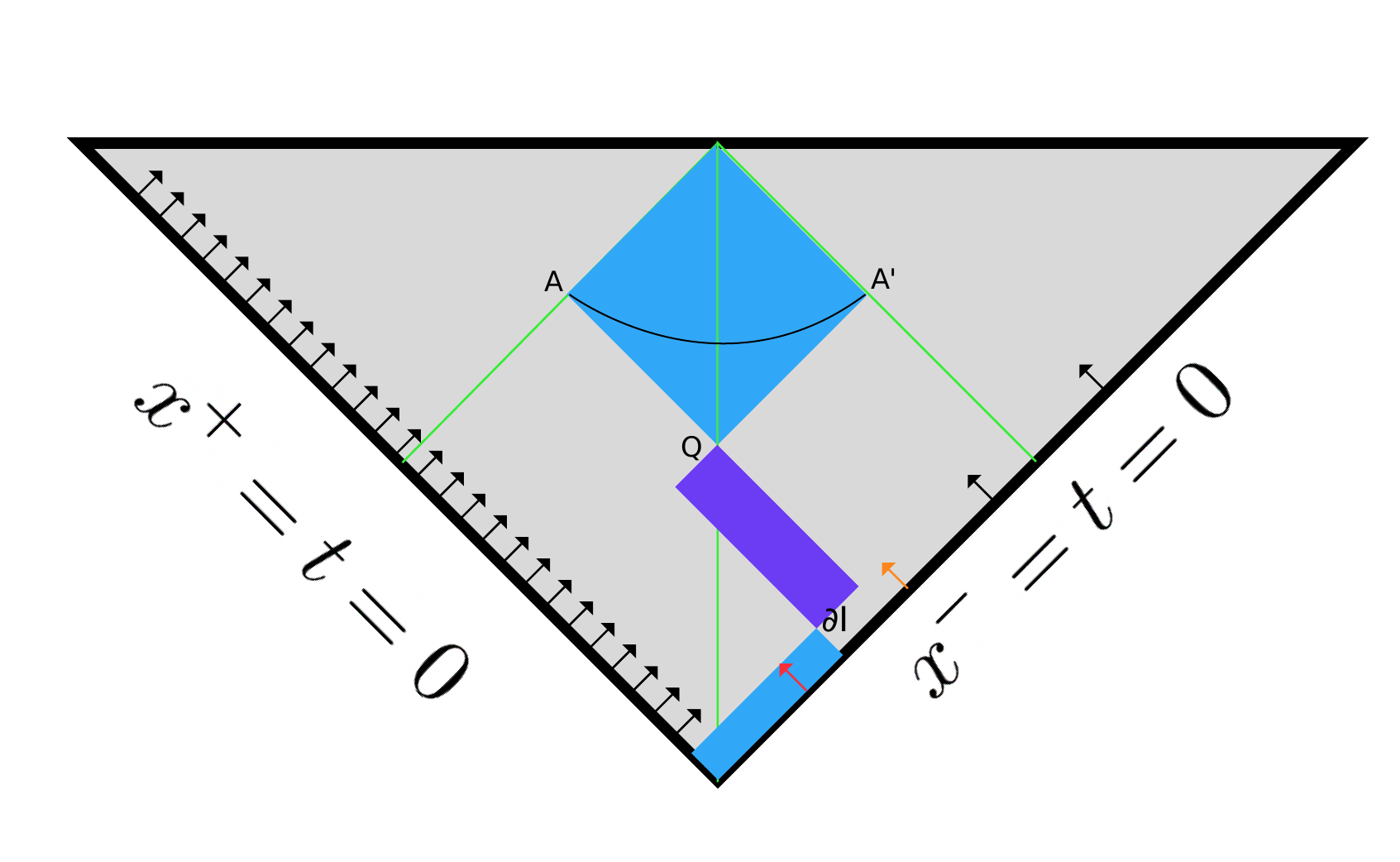}
\centering
\caption{The domain of dependence of the radiation between the 2 horizons is the upper blue region. The island is the lower blue region. The left horizon is evaporating but the right horizon is growing in this non-equilibrium state. Note that the red particle will go inside the left horizon before the orange particle. Because of that, we can get information about the red particle before the orange particle using Hawking radiation. This is reflected in the island covering the red particle even before it covers the orange particle. The timelike entanglement entropy for the union of CFT in blue regions is equal to the value for the purple region.}
\label{fig:island}
\end{figure}

In Fig \ref{fig:island}, the radiation between the horizons is in between $A$ and $A'$, where $A$ and $A'$ are time-dependent and are moving along the horizons. They are given by

$$(x_{A}^{+},x_{A}^{-})=(\tanh(t/2)e^{-\sqrt{\dfrac{2}{\sinh (2t)}}},\tanh(t/2)e^{+\sqrt{\dfrac{2}{\sinh (2t)}}})$$
$$(x_{A'}^{+},x_{A'}^{-})=(\tanh(t/2)e^{+\sqrt{\dfrac{2}{\sinh (2t)}}},\tanh(t/2)e^{-\sqrt{\dfrac{2}{\sinh (2t)}}})$$

To find $S_{\mathrm{vN}}(R \cup I)$, we should find the entropy of the interval between the points $\partial I$ and $Q$. Here, $Q=(x_Q^+,x_Q^-)=(x_A^+,x_{A'}^-)$ as shown in Fig \ref{fig:island}. The von Neumann entropy for the union of CFT in $R$ and $I$ is equal to the value of the von Neumann entropy for the interval between the points $\partial I$ and $Q$. Note that when calculating time like entanglement we take points that are above and below instead of the usual case where we take left and right.
$$
\begin{aligned}
S_{\mathrm{vN}}(R \cup I)&=\frac{c}{12} \log \left[\frac{4 \left(x_{\partial I}^{+} x_Q^{+}\right)^{1 / 2}\left(x_{\partial I}^{-}-x_Q^{-}\right)^2}{\epsilon^2\left(1-x_{\partial I}^{+} x_{\partial I}^{-}\right)\left(1-x_Q^{+} x_Q^{-}\right)}\right] \\
&+\frac{c}{12} \log \left[\frac{4\left(x_{\partial I}^{+} x_Q^{+}\right)^{1 / 2} \log \left(x_Q^{+} / x_{\partial I}^{+}\right)^2}{\epsilon^2\left(1-x_{\partial I}^{+} x_{\partial I}^{-}\right)\left(1-x_Q^{+} x_Q^{-}\right)}\right]
\end{aligned}
$$
with
$$(x_{Q}^{+},x_{Q}^{-})=(\tanh(t/2)e^{-\sqrt{\dfrac{2}{\sinh (2t)}}},\tanh(t/2)e^{-\sqrt{\dfrac{2}{\sinh (2t)}}})$$
In appendix \ref{appendix:QES}, using the above equation and equation \ref{eq:Phi}, we solve for the island assuming late times (since island contribution happens only at late times) and obtain 

$$(x_{\partial I}^{+},x_{\partial I}^{-}) \approx \left(\frac{-2}{x_Q^{-}W_0\left(\frac{-2}{x_Q^{-}}e^{-2-\dfrac{2 \Phi_s}{c}}\right)},\frac{1}{x_{\partial I}^{+}\log \left(x_{\partial I}^{+}\right)}\right)$$

This solution passes some intuitive checks. As shown in the Fig \ref{fig:island}, particles\footnote{In CFTs, there is no proper particle interpretation for excitations. So this usage is just for the sake of intuition and explanation.} which go into the left horizon first are recovered from the Hawking radiation first. $x_{\partial I}^{+}$ increases with time and $x_{\partial I}^{-}$ decreases with time.

\subsection{Page curve}

If we now consider the \textit{no island case} and calculate the von Neumann entropy between points $A$ and $A'$ by substituting the points we obtain

$$
\begin{aligned}
S_{\mathrm{vN}}(R) &=\frac{c}{12} \log \left[\frac{4 \left(x_A^{+} x_{A^{\prime}}^{+}\right)\left(x_A^{-}-x_{A^{\prime}}^{-}\right)^2\log \left(x_{A^{\prime}}^{+} / x_A^{+}\right)^2}{\epsilon^4\left(1-x_A^{+} x_A^{-}\right)^2\left(1-x_{A^{\prime}}^{+} x_{A^{\prime}}^{-}\right)^2}\right]\\
&=\frac{c}{12} \log \left[\sinh^4 (t)\right]+\text{some constant}\\
&\approx\frac{c}{12} \log \left[\left(\frac{e^t}{2}\right)^4\right]+\text{some constant}\\
&\approx\frac{c}{12} (4t)+\text{some constant}.
\end{aligned}
$$
This is a strictly increasing function, as expected.

Now for the \textit{island case}, we can use an expansion for the Lambert W function expansion and plot the final equation in any software, and we can get that for large $t$

$$
\begin{aligned}
S_{\mathrm{vN}}(R \cup I)&=\frac{c}{12} \log \left[\frac{16 \left(x_{\partial I}^{+} x_Q^{+}\right)\left(x_{\partial I}^{-}-x_Q^{-}\right)^2\log \left(x_Q^{+} / x_{\partial I}^{+}\right)^2}{\epsilon^4\left(1-x_{\partial I}^{+} x_{\partial I}^{-}\right)^2\left(1-x_Q^{+} x_Q^{-}\right)^2}\right] \\
&\approx \frac{c}{12} (0.8t) +\text{some constant}.
\end{aligned}
$$
At first glance, this might seem opposite to our expectations from seeing the black hole case. We generally expect that because of the island the entropy will start decreasing after the Page time. But here, we have to note that \textbf{another competing phenomenon} is going on. Remember that once Hawking radiation is emitted from the left horizon after some time, it will also go beyond the right horizon. So, some Hawking radiation that went beyond the right horizon will be entangled with the Hawking radiation that is between the horizons. This will obviously increase the entropy of the radiation between the horizons, i.e., radiation between points $A$ and $A'$. The other phenomenon is due to the island, which will decrease the entropy of the radiation. What we are seeing is the net phenomenon. We can interpret the result as saying that the increasing phenomenon dominates slightly over the decreasing phenomenon.

Another thing to keep in mind is that the Page curve might be only valid for the case of spacelike entanglement, and for the timelike case, \textbf{we might need to generalize the idea of the Page curve to something broader}.

\section{Discussion}

In this paper, we showed that even in FLRW cosmologies with positive cosmological constant and with no spatial curvature, islands can exist once we consider timelike entanglement. Fig \ref{fig:island} summarizes the main result of this paper. Our results suggest that the information about the particles that went beyond the horizon is not truly lost. By measuring Hawking radiation, we can obtain information about the particles that went beyond the horizon directly from the past when were near the initial singularity. This is similar to how black holes violate locality, but here causality is violated. Our island solution passes intuitive checks like the particles that went into the horizon first can be accessed first by measuring Hawking radiation. By measuring Hawking radiation, we will not get information about the present particles that are beyond the horizon. Instead, we get the same information from the past when these particles were still inside the observable universe of our observer.

Recall that black holes violate locality because, after the Page time, the island and the radiation become entangled. However, particles generally get entangled locally at the same spacetime region, and then the entanglement will be present even if they are separated. \textit{In the normal island's case, even though the radiation and the island have a large spatial separation, entanglement is forming between them}. This is \textbf{nonlocal} behavior. In our case, the timelike entanglement between the island and the radiation forms after the Page time. There is a large timelike separation between the island and the radiation when timelike entanglement forms. So, this is \textbf{noncausal} behavior.

But, in this paper, we assumed that the island prescription is valid even when the island is timelike entangled. In future work, we may try to derive this generalized island prescription that includes timelike entanglement, either using dS/CFT correspondence (once it is understood) or replica wormholes in the gravitational path integral approach.

In $2D$, there are two disjoint horizons, but in higher dimensions, there will be a single connected horizon. Because of this, there might be some differences in higher dimensions. It would also be interesting to study timelike entangled islands in higher dimensions.

Based on our results, we propose a natural generalization of the Ryu–Takayanagi formula in $dS_2$ holography. The generalization will be nontrivial for higher dimensional de Sitter spacetimes, and we have not yet figured it out.

\subsection{Conjecture: Ryu–Takayanagi or QES formula in $dS_2$ holography}
\begin{figure}[h]
\includegraphics[width=8cm]{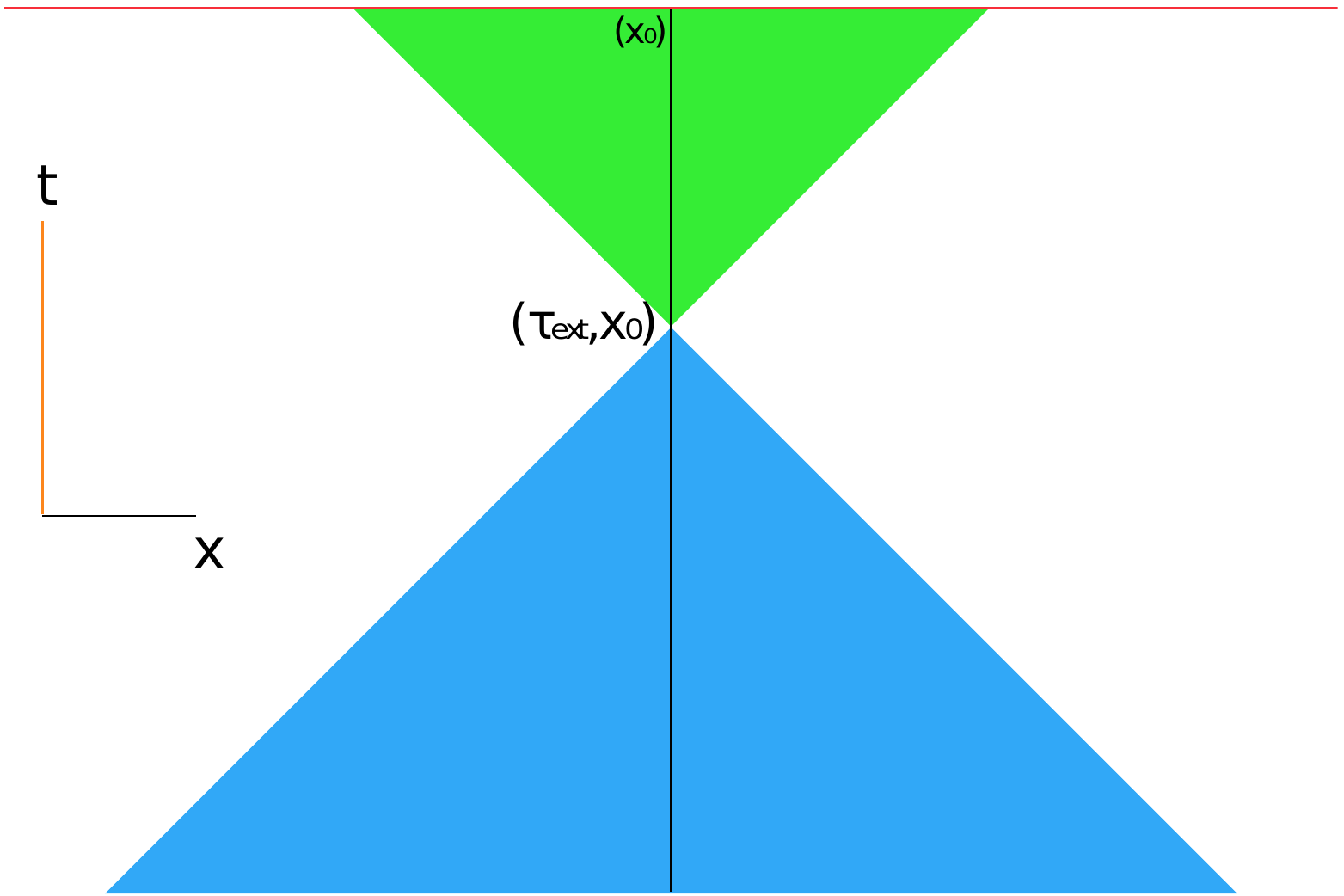}
\centering
\caption{The \textbf{timelike} entanglement between the green region and the blue region will be equal to the \textbf{spacelike} entanglement between the boundary regions $x<x_0$ and $x>x_0$. The boundary theory lives on the spatial line shown in red.}
\label{fig:global}
\end{figure}

Take the global solution for the de Sitter JT gravity
$$
d s_{\text {Global }}^2=-d t^2+\cosh ^2 (t) d x^2, \quad \Phi=\Phi_0 \sinh t .
$$
Now the boundary theory dual to this bulk gravitational theory will live on the future boundary ($t\to \infty$). As is well known, in the dS/CFT Correspondence, the dual theory will not have any timelike dimensions. Now take a point $x_0$ in the dual theory. This point divides the boundary into two subregions, $x<x_0$ and $x>x_0$. The entanglement entropy between these two boundary subregions will be given by
$$S=\min \operatorname{ext}_{\tau}\left[\frac{\Phi(\tau,x_0)}{4 G_N}\right].$$
This is the generalization of the Ryu–Takayanagi formula in $dS_2$ holography. The QES formula will then be the following generalization of the above formula
$$S=\min \operatorname{ext}_{\tau}\left[\frac{\Phi(\tau,x_0)}{4 G_N}+S_{\mathrm{vN}}(t>\tau,x_0)\right].$$
Here $S_{\mathrm{vN}}(t>\tau,x_0)$ is the timelike entanglement entropy of the domain of dependence that will be uniquely defined by the timelike semi-infinite interval $(t>\tau,x_0)$. The quantum extremal surface $(\tau_{ext},x_0)$ shown in Fig \ref{fig:global} will minimize the generalized entropy.

Note that the \textbf{bulk} entanglement we are talking about is \textbf{timelike}, but the \textbf{boundary} entanglement we are considering is \textbf{spacelike}. In the AdS/CFT correspondence, the emergent dimension is a spatial dimension, but here the \textbf{emergent dimension is a temporal dimension}. This is reflected in the fact that the RT or QES formula now gives the bulk timelike entanglement. We believe that timelike entanglement will become as crucial in dS/CFT as spacelike entanglement is in AdS/CFT. So, more understanding about timelike entanglement is needed.

\appendix
\section{Calculation of QES}
\label{appendix:QES}
Let $2 \Phi_H(\partial I)+S_{\mathrm{vN}}(R \cup I)=\dfrac{c}{24}f(\partial I)$ then $S(R)=\min \operatorname{ext}_I\left[\dfrac{c}{48}f(\partial I)\right]$\footnote{Full calculation available at \url{https://ksr.onl/papers/1/QEScalculation.pdf}}.

$$
\begin{aligned}
\dfrac{c}{24}f(\partial I)&=\frac{c}{24}\left[1+\frac{2 \Phi_s}{c} \left(\frac{1+x_{\partial I}^{+}x_{\partial I}^{-}}{1-x_{\partial I}^{+}x_{\partial I}^{-}}\right)- \left(\frac{1+x_{\partial I}^{+}x_{\partial I}^{-}}{1-x_{\partial I}^{+}x_{\partial I}^{-}}\right) \log \left(x_{\partial I}^{+}\right)\right]\\
 &+\frac{c}{12} \log \left[\frac{4 \left(x_{\partial I}^{+} x_Q^{+}\right)^{1 / 2}\left(x_{\partial I}^{-}-x_Q^{-}\right)^2}{\epsilon^2\left(1-x_{\partial I}^{+} x_{\partial I}^{-}\right)\left(1-x_Q^{+} x_Q^{-}\right)}\right] \\
&+\frac{c}{12} \log \left[\frac{4\left(x_{\partial I}^{+} x_Q^{+}\right)^{1 / 2} \log \left(x_Q^{+} / x_{\partial I}^{+}\right)^2}{\epsilon^2\left(1-x_{\partial I}^{+} x_{\partial I}^{-}\right)\left(1-x_Q^{+} x_Q^{-}\right)}\right]
\end{aligned}
$$
Now to find the minima of $f(\partial I)$ we partial differentiate it with respect to $x_{\partial I}^{+}$ and $x_{\partial I}^{-}$.

Partial differentiation with respect to $x_{\partial I}^{+}$ gives

$$
\boxed{\implies\begin{aligned}
0&=\left[\frac{2 \Phi_s}{c} \left(\frac{2x_{\partial I}^{-}}{(1-x_{\partial I}^{+}x_{\partial I}^{-})^2}\right)-\left(\frac{2x_{\partial I}^{-}}{(1-x_{\partial I}^{+}x_{\partial I}^{-})^2}\right)\log \left(x_{\partial I}^{+}\right)- \left(\frac{1+x_{\partial I}^{+}x_{\partial I}^{-}}{1-x_{\partial I}^{+}x_{\partial I}^{-}}\right) \frac{1}{x_{\partial I}^{+}}\right]\\
&+2 \left[\frac{1}{2x_{\partial I}^{+}}+\frac{x_{\partial I}^{-}}{\left(1-x_{\partial I}^{+} x_{\partial I}^{-}\right)}\right] \\
&+2 \left[\frac{1}{2x_{\partial I}^{+}}+\frac{x_{\partial I}^{-}}{\left(1-x_{\partial I}^{+} x_{\partial I}^{-}\right)}-\frac{2}{x_{\partial I}^{+}\log \left(x_Q^{+} / x_{\partial I}^{+}\right)}\right]\\
\end{aligned}}
$$

Similarly with respect to $x_{\partial I}^{-}$ gives

$$
\boxed{\implies\begin{aligned}
0&=\left(\frac{2x_{\partial I}^{+}}{(1-x_{\partial I}^{+}x_{\partial I}^{-})^2}\right)\left(\frac{2 \Phi_s}{c}-\log \left(x_{\partial I}^{+}\right)\right) \\
 &+2 \left[2\frac{1}{\left(x_{\partial I}^{-}-x_Q^{-}\right)}+\frac{x_{\partial I}^{+}}{\left(1-x_{\partial I}^{+} x_{\partial I}^{-}\right)}\right] \\
&+2 \frac{x_{\partial I}^{+}}{\left(1-x_{\partial I}^{+} x_{\partial I}^{-}\right)}
\end{aligned}}
$$

Now in the above equation if we assume $x_{\partial I}^{-}<< 1$

$$
\implies\begin{aligned}
0&=2x_{\partial I}^{+}\left(\frac{2 \Phi_s}{c}-\log \left(x_{\partial I}^{+}\right)\right) \\
 &+2 \left[-2\frac{1}{x_Q^{-}}+x_{\partial I}^{+}\right] \\
&+2 x_{\partial I}^{+}
\end{aligned}
$$
then

$$\boxed{x_{\partial I}^{+}=\frac{-2}{x_Q^{-}W_0\left(\frac{-2}{x_Q^{-}}e^{-2-\dfrac{2 \Phi_s}{c}}\right)}}$$

Now if we take the 1st boxed equation and assume $x_{\partial I}^{-}<< 1$ it gives

$$
\implies\begin{aligned}
0&=\left[\left(\frac{2 \Phi_s}{c}-\log \left(x_{\partial I}^{+}\right)\right)2x_{\partial I}^{-}- \frac{1}{x_{\partial I}^{+}}\right]\\
&+2 \left[\frac{1}{2x_{\partial I}^{+}}+x_{\partial I}^{-}\right] \\
&+2 \left[\frac{1}{2x_{\partial I}^{+}}+x_{\partial I}^{-}-\frac{2}{x_{\partial I}^{+}\log \left(x_Q^{+} / x_{\partial I}^{+}\right)}\right]\\
\end{aligned}
$$

This gives us the approximation at late times
$$\boxed{x_{\partial I}^{-} \approx \frac{1}{x_{\partial I}^{+}\log \left(x_{\partial I}^{+}\right)}<<1}$$

\acknowledgments

I want to thank Pichai Ramadevi for the helpful discussions.

\noindent I also want to thank Einstein, Newton, Gautama, and Goku for inspiring me since childhood.

\paragraph{Note added.} Originally I conjectured the formula for entanglement entropy of a timelike interval in 2D CFT \ref{eq:timelikeinterval} by giving some intuitive reasons. After almost completing this paper, I noticed the preprint \cite{Doi:2022iyj} which derives the same formula. I removed it as a conjecture and used their result by citing them.


\end{document}